

\magnification=\magstep1


\newbox\SlashedBox
\def\slashed#1{\setbox\SlashedBox=\hbox{#1}
\hbox to 0pt{\hbox to 1\wd\SlashedBox{\hfil/\hfil}\hss}#1}
\def\hboxtosizeof#1#2{\setbox\SlashedBox=\hbox{#1}
\hbox to 1\wd\SlashedBox{#2}}

\def\mathslashed#1{\setbox\SlashedBox=\hbox{$#1$}
\hbox to 0pt{\hbox to 1\wd\SlashedBox{\hfil/\hfil}\hss}#1}

\def\ifsmall{\iffalse}  
\def\titlepagefont{}  

\def\DefineTeXgraphics{%
\special{ps::[global] /TeXgraphics { } def}}  

\def\today{\ifcase\month\or January\or February\or March\or April\or May
\or June\or July\or August\or September\or October\or November\or
December\fi\space\number\day, \number\year}
\def\eatPrefix19{}
\def\Year{\expandafter\eatPrefix\the\year}
\newcount\hours \newcount\minutes
\def\monthname{\ifcase\month\or
January\or February\or March\or April\or May\or June\or July\or
August\or September\or October\or November\or December\fi}
\def\shortmonthname{\ifcase\month\or
Jan\or Feb\or Mar\or Apr\or May\or Jun\or Jul\or
Aug\or Sep\or Oct\or Nov\or Dec\fi}

\def\TimeStamp{\hours\the\time\divide\hours by60%
\minutes -\the\time\divide\minutes by60\multiply\minutes by60%
\advance\minutes by\the\time%
${\rm \shortmonthname}\cdot\if\day<10{}0\fi\the\day\cdot\the\year%
\qquad\the\hours:\if\minutes<10{}0\fi\the\minutes$}




\def\Title#1{%
\vskip 1in{\titlefont\centerline{#1}}\vskip .5in}



\newif\ifdraftmode

\def\nolabels{\def\wrlabeL##1{}\def\eqlabeL##1{}\def\reflabeL##1{}}
\def\writelabels{\def\wrlabeL##1{\leavevmode\vadjust{\rlap{\smash%
{\line{{\escapechar=` \hfill\rlap{\sevenrm\hskip.03in\string##1}}}}}}}%
\def\eqlabeL##1{{\escapechar-1\rlap{\sevenrm\hskip.05in\string##1}}}%
\def\reflabeL##1{\noexpand\llap{\noexpand\sevenrm\string\string\string##1}}}
\nolabels

\newdimen\fullhsize
\newdimen\hstitle
\hstitle=\hsize 
\newdimen\hsbody
\hsbody=\hsize 
\newdimen\hbodyoffset
\hbodyoffset=\hoffset 
\newbox\leftpage
\def\abstract#1{#1}
\def\rotated{\special{ps: landscape}
\magnification=1000  
\baselineskip=14pt
\global\hstitle=9truein\global\hsbody=4.75truein
\global\vsize=7truein\global\voffset=-.31truein
\global\hoffset=-0.54in\global\hbodyoffset=-.54truein
\global\fullhsize=10truein
\def\DefineTeXgraphics{%
\special{ps::[global]
/TeXgraphics {currentpoint translate 0.7 0.7 scale
              -80 0.72 mul -1000 0.72 mul translate} def}}
\let\lr=L
\def\ifsmall{\iftrue}
\def\titlepagefont{\twelvepoint}
\trueseventeenpoint
\def\almostshipout##1{\if L\lr \count1=1
      \global\setbox\leftpage=##1 \global\let\lr=R
   \else \count1=2
      \shipout\vbox{\hbox to\fullhsize{\box\leftpage\hfil##1}}
      \global\let\lr=L\fi}

\output={\ifnum\count0=1 
 \shipout\vbox{\hbox to \fullhsize{\hfill\pagebody\hfill}}\advancepageno
 \else
 \almostshipout{\leftline{\vbox{\pagebody\makefootline}}}\advancepageno
 \fi}

\def\abstract##1{{\leftskip=1.5in\rightskip=1.5in ##1\par}} }

\def\linemessage#1{\immediate\write16{#1}}

\global\newcount\secno \global\secno=0
\global\newcount\appno \global\appno=0
\global\newcount\meqno \global\meqno=1
\global\newcount\subsecno \global\subsecno=0
\global\newcount\figno \global\figno=0

\newif\ifAnyCounterChanged
\let\terminator=\relax
\def\normalize#1{\ifx#1\terminator\let\next=\relax\else%
\if#1i\aftergroup i\else\if#1v\aftergroup v\else\if#1x\aftergroup x%
\else\if#1l\aftergroup l\else\if#1c\aftergroup c\else%
\if#1m\aftergroup m\else%
\if#1I\aftergroup I\else\if#1V\aftergroup V\else\if#1X\aftergroup X%
\else\if#1L\aftergroup L\else\if#1C\aftergroup C\else%
\if#1M\aftergroup M\else\aftergroup#1\fi\fi\fi\fi\fi\fi\fi\fi\fi\fi\fi\fi%
\let\next=\normalize\fi%
\next}
\def\makeNormal#1#2{\def\doNormalDef{\edef#1}\begingroup%
\aftergroup\doNormalDef\aftergroup{\normalize#2\terminator\aftergroup}%
\endgroup}

\def\warnIfChanged#1#2{%
\ifundef#1
\else\begingroup%
\edef\oldDefinitionOfCounter{#1}\edef\newDefinitionOfCounter{#2}%
\ifx\oldDefinitionOfCounter\newDefinitionOfCounter%
\else%
\linemessage{Warning: definition of \noexpand#1 has changed.}%
\global\AnyCounterChangedtrue\fi\endgroup\fi}

\def\Section#1{\global\advance\secno by1\relax\global\meqno=1%
\global\subsecno=0%
\bigbreak\bigskip
\centerline{\twelvepoint \bf %
\the\secno. #1}%
\par\nobreak\medskip\nobreak}
\def\tagsection#1{%
\warnIfChanged#1{\the\secno}%
\xdef#1{\the\secno}%
\ifWritingAuxFile\immediate\write\auxfile{\noexpand\xdef\noexpand#1{#1}}\fi%
}
\def\section{\Section}
\def\Subsection#1{\global\advance\subsecno by1\relax\medskip %
\leftline{\bf\the\secno.\the\subsecno\ #1}%
\par\nobreak\smallskip\nobreak}
\def\tagsubsection#1{%
\warnIfChanged#1{\the\secno.\the\subsecno}%
\xdef#1{\the\secno.\the\subsecno}%
\ifWritingAuxFile\immediate\write\auxfile{\noexpand\xdef\noexpand#1{#1}}\fi%
}

\def\subsection{\Subsection}

\def\romappno{\uppercase\expandafter{\romannumeral\appno}}
\def\makeNormalizedRomappno{%
\expandafter\makeNormal\expandafter\normalizedromappno%
\expandafter{\romannumeral\appno}%
\edef\normalizedromappno{\uppercase{\normalizedromappno}}}
\def\Appendix#1{\global\advance\appno by1\relax\global\meqno=1\global\secno=0
\bigbreak\bigskip
\centerline{\twelvepoint \bf Appendix %
\romappno. #1}%
\par\nobreak\medskip\nobreak}
\def\tagappendix#1{\makeNormalizedRomappno%
\warnIfChanged#1{\normalizedromappno}%
\xdef#1{\normalizedromappno}%
\ifWritingAuxFile\immediate\write\auxfile{\noexpand\xdef\noexpand#1{#1}}\fi%
}
\def\appendix{\Appendix}

\def\eqn#1{\makeNormalizedRomappno%
\ifnum\secno>0%
  \warnIfChanged#1{\the\secno.\the\meqno}%
  \eqno(\the\secno.\the\meqno)\xdef#1{\the\secno.\the\meqno}%
     \global\advance\meqno by1
\else\ifnum\appno>0%
  \warnIfChanged#1{\normalizedromappno.\the\meqno}%
  \eqno({\rm\romappno}.\the\meqno)%
      \xdef#1{\normalizedromappno.\the\meqno}%
     \global\advance\meqno by1
\else%
  \warnIfChanged#1{\the\meqno}%
  \eqno(\the\meqno)\xdef#1{\the\meqno}%
     \global\advance\meqno by1
\fi\fi%
\eqlabeL#1%
\ifWritingAuxFile\immediate\write\auxfile{\noexpand\xdef\noexpand#1{#1}}\fi%
}
\def\defeqn#1{\makeNormalizedRomappno%
\ifnum\secno>0%
  \warnIfChanged#1{\the\secno.\the\meqno}%
  \xdef#1{\the\secno.\the\meqno}%
     \global\advance\meqno by1
\else\ifnum\appno>0%
  \warnIfChanged#1{\normalizedromappno.\the\meqno}%
  \xdef#1{\normalizedromappno.\the\meqno}%
     \global\advance\meqno by1
\else%
  \warnIfChanged#1{\the\meqno}%
  \xdef#1{\the\meqno}%
     \global\advance\meqno by1
\fi\fi%
\eqlabeL#1%
\ifWritingAuxFile\immediate\write\auxfile{\noexpand\xdef\noexpand#1{#1}}\fi%
}
\def\anoneqn{\makeNormalizedRomappno%
\ifnum\secno>0
  \eqno(\the\secno.\the\meqno)%
     \global\advance\meqno by1
\else\ifnum\appno>0
  \eqno({\rm\normalizedromappno}.\the\meqno)%
     \global\advance\meqno by1
\else
  \eqno(\the\meqno)%
     \global\advance\meqno by1
\fi\fi%
}
\def\mfig#1#2{\global\advance\figno by1%
\relax#1\the\figno%
\warnIfChanged#2{\the\figno}%
\edef#2{\the\figno}%
\reflabeL#1%
\ifWritingAuxFile\immediate\write\auxfile{\noexpand\xdef\noexpand#2{#2}}\fi%
}

\catcode`@=11 

\font\ninerm=cmr9
\font\eightrm=cmr8
\font\sixrm=cmr6

\def\loadtrueseventeenpoint{
 \font\seventeenrm=cmr10 at 17.28truept
 \font\seventeeni=cmmi10 at 17.28truept
 \font\seventeenbf=cmbx10 at 17.28truept
 \font\seventeenit=cmti10 at 17.28truept
 \font\seventeensl=cmsl10 at 17.28truept
 \font\seventeensy=cmsy10 at 17.28truept
}
\def\loadfourteenpoint{
\font\fourteenrm=cmr10 at 14.4pt
\font\fourteeni=cmmi10 at 14.4pt
\font\fourteenit=cmti10 at 14.4pt
\font\fourteensl=cmsl10 at 14.4pt
\font\fourteensy=cmsy10 at 14.4pt
\font\fourteenbf=cmbx10 at 14.4pt
}
\def\loadtruetwelvepoint{
\font\twelverm=cmr10 at 12truept
\font\twelvei=cmmi10 at 12truept
\font\twelveit=cmti10 at 12truept
\font\twelvesl=cmsl10 at 12truept
\font\twelvesy=cmsy10 at 12truept
\font\twelvebf=cmbx10 at 12truept
}

\font\ninei=cmmi9
\font\eighti=cmmi8
\font\sixi=cmmi6
\skewchar\ninei='177 \skewchar\eighti='177 \skewchar\sixi='177

\font\ninesy=cmsy9
\font\eightsy=cmsy8
\font\sixsy=cmsy6
\skewchar\ninesy='60 \skewchar\eightsy='60 \skewchar\sixsy='60

\font\ninebf=cmbx9
\font\eightbf=cmbx8
\font\sixbf=cmbx6

\font\ninett=cmtt9
\font\eighttt=cmtt8

\hyphenchar\tentt=-1 
\hyphenchar\ninett=-1
\hyphenchar\eighttt=-1

\font\ninesl=cmsl9
\font\eightsl=cmsl8

\font\nineit=cmti9
\font\eightit=cmti8


\newskip\ttglue
\def\tenpoint{\def\rm{\fam0\tenrm}%
  \textfont0=\tenrm \scriptfont0=\sevenrm \scriptscriptfont0=\fiverm
  \textfont1=\teni \scriptfont1=\seveni \scriptscriptfont1=\fivei
  \textfont2=\tensy \scriptfont2=\sevensy \scriptscriptfont2=\fivesy
  \textfont3=\tenex \scriptfont3=\tenex \scriptscriptfont3=\tenex
  \def\it{\fam\itfam\tenit}\textfont\itfam=\tenit
  \def\sl{\fam\slfam\tensl}\textfont\slfam=\tensl
  \def\bf{\fam\bffam\tenbf}\textfont\bffam=\tenbf \scriptfont\bffam=\sevenbf
  \scriptscriptfont\bffam=\fivebf
  \normalbaselineskip=12pt
  \let\sc=\eightrm
  \let\big=\tenbig
  \setbox\strutbox=\hbox{\vrule height8.5pt depth3.5pt width\z@}%
  \normalbaselines\rm}

\def\twelvepoint{\def\rm{\fam0\twelverm}%
  \textfont0=\twelverm \scriptfont0=\ninerm \scriptscriptfont0=\sevenrm
  \textfont1=\twelvei \scriptfont1=\ninei \scriptscriptfont1=\seveni
  \textfont2=\twelvesy \scriptfont2=\ninesy \scriptscriptfont2=\sevensy
  \textfont3=\tenex \scriptfont3=\tenex \scriptscriptfont3=\tenex
  \def\it{\fam\itfam\twelveit}\textfont\itfam=\twelveit
  \def\sl{\fam\slfam\twelvesl}\textfont\slfam=\twelvesl
  \def\bf{\fam\bffam\twelvebf}\textfont\bffam=\twelvebf
\scriptfont\bffam=\nineb
   f
  \scriptscriptfont\bffam=\sevenbf
  \normalbaselineskip=12pt
  \let\sc=\eightrm
  \let\big=\tenbig
  \setbox\strutbox=\hbox{\vrule height8.5pt depth3.5pt width\z@}%
  \normalbaselines\rm}

\def\fourteenpoint{\def\rm{\fam0\fourteenrm}%
  \textfont0=\fourteenrm \scriptfont0=\tenrm \scriptscriptfont0=\sevenrm
  \textfont1=\fourteeni \scriptfont1=\teni \scriptscriptfont1=\seveni
  \textfont2=\fourteensy \scriptfont2=\tensy \scriptscriptfont2=\sevensy
  \textfont3=\tenex \scriptfont3=\tenex \scriptscriptfont3=\tenex
  \def\it{\fam\itfam\fourteenit}\textfont\itfam=\fourteenit
  \def\sl{\fam\slfam\fourteensl}\textfont\slfam=\fourteensl
  \def\bf{\fam\bffam\fourteenbf}\textfont\bffam=\fourteenbf%
  \scriptfont\bffam=\tenbf
  \scriptscriptfont\bffam=\sevenbf
  \normalbaselineskip=17pt
  \let\sc=\elevenrm
  \let\big=\tenbig
  \setbox\strutbox=\hbox{\vrule height8.5pt depth3.5pt width\z@}%
  \normalbaselines\rm}

\def\seventeenpoint{\def\rm{\fam0\seventeenrm}%
  \textfont0=\seventeenrm \scriptfont0=\fourteenrm \scriptscriptfont0=\tenrm
  \textfont1=\seventeeni \scriptfont1=\fourteeni \scriptscriptfont1=\teni
  \textfont2=\seventeensy \scriptfont2=\fourteensy \scriptscriptfont2=\tensy
  \textfont3=\tenex \scriptfont3=\tenex \scriptscriptfont3=\tenex
  \def\it{\fam\itfam\seventeenit}\textfont\itfam=\seventeenit
  \def\sl{\fam\slfam\seventeensl}\textfont\slfam=\seventeensl
  \def\bf{\fam\bffam\seventeenbf}\textfont\bffam=\seventeenbf%
  \scriptfont\bffam=\fourteenbf
  \scriptscriptfont\bffam=\twelvebf
  \normalbaselineskip=21pt
  \let\sc=\fourteenrm
  \let\big=\tenbig
  \setbox\strutbox=\hbox{\vrule height 12pt depth 6pt width\z@}%
  \normalbaselines\rm}

\def\ninepoint{\def\rm{\fam0\ninerm}%
  \textfont0=\ninerm \scriptfont0=\sixrm \scriptscriptfont0=\fiverm
  \textfont1=\ninei \scriptfont1=\sixi \scriptscriptfont1=\fivei
  \textfont2=\ninesy \scriptfont2=\sixsy \scriptscriptfont2=\fivesy
  \textfont3=\tenex \scriptfont3=\tenex \scriptscriptfont3=\tenex
  \def\it{\fam\itfam\nineit}\textfont\itfam=\nineit
  \def\sl{\fam\slfam\ninesl}\textfont\slfam=\ninesl
  \def\bf{\fam\bffam\ninebf}\textfont\bffam=\ninebf \scriptfont\bffam=\sixbf
  \scriptscriptfont\bffam=\fivebf
  \normalbaselineskip=11pt
  \let\sc=\sevenrm
  \let\big=\ninebig
  \setbox\strutbox=\hbox{\vrule height8pt depth3pt width\z@}%
  \normalbaselines\rm}

\def\eightpoint{\def\rm{\fam0\eightrm}%
  \textfont0=\eightrm \scriptfont0=\sixrm \scriptscriptfont0=\fiverm%
  \textfont1=\eighti \scriptfont1=\sixi \scriptscriptfont1=\fivei%
  \textfont2=\eightsy \scriptfont2=\sixsy \scriptscriptfont2=\fivesy%
  \textfont3=\tenex \scriptfont3=\tenex \scriptscriptfont3=\tenex%
  \def\it{\fam\itfam\eightit}\textfont\itfam=\eightit%
  \def\sl{\fam\slfam\eightsl}\textfont\slfam=\eightsl%
  \def\bf{\fam\bffam\eightbf}\textfont\bffam=\eightbf \scriptfont\bffam=\sixbf%
  \scriptscriptfont\bffam=\fivebf%
  \normalbaselineskip=9pt%
  \let\sc=\sixrm%
  \let\big=\eightbig%
  \setbox\strutbox=\hbox{\vrule height7pt depth2pt width\z@}%
  \normalbaselines\rm}

\def\tenbig#1{{\hbox{$\left#1\vbox to8.5pt{}\right.\n@space$}}}
\def\ninebig#1{{\hbox{$\textfont0=\tenrm\textfont2=\tensy
  \left#1\vbox to7.25pt{}\right.\n@space$}}}
\def\eightbig#1{{\hbox{$\textfont0=\ninerm\textfont2=\ninesy
  \left#1\vbox to6.5pt{}\right.\n@space$}}}

\def\footnote#1{\edef\@sf{\spacefactor\the\spacefactor}#1\@sf
      \insert\footins\bgroup\eightpoint
      \interlinepenalty100 \let\par=\endgraf
        \leftskip=\z@skip \rightskip=\z@skip
        \splittopskip=10pt plus 1pt minus 1pt \floatingpenalty=20000
        \smallskip\item{#1}\bgroup\strut\aftergroup\@foot\let\next}
\skip\footins=12pt plus 2pt minus 4pt 
\dimen\footins=30pc 

\newinsert\margin
\dimen\margin=\maxdimen
\def\titlefont{\seventeenpoint}
\loadtruetwelvepoint 
\loadtrueseventeenpoint
\catcode`\@=\active
\catcode`@=12  
\catcode`\"=\active

\def\eatOne#1{}
\def\ifundef#1{\expandafter\ifx%
\csname\expandafter\eatOne\string#1\endcsname\relax}
\def\notTrue{\iffalse}\def\isTrue{\iftrue}
\def\ifdef#1{{\ifundef#1%
\aftergroup\notTrue\else\aftergroup\isTrue\fi}}
\def\use#1{\ifundef#1\linemessage{Warning: \string#1 is undefined.}%
{\tt \string#1}\else#1\fi}


\global\newcount\refno \global\refno=1
\newwrite\rfile
\newlinechar=`\^^J
\def\ref#1#2{\the\refno\nref#1{#2}}
\def\nref#1#2{\xdef#1{\the\refno}%
\ifnum\refno=1\immediate\openout\rfile=refs.tmp\fi%
\immediate\write\rfile{\noexpand\item{[\noexpand#1]\ }#2.}%
\global\advance\refno by1}
\def\lref#1#2{\the\refno\xdef#1{\the\refno}%
\ifnum\refno=1\immediate\openout\rfile=refs.tmp\fi%
\immediate\write\rfile{\noexpand\item{[\noexpand#1]\ }#2\semi}%
\global\advance\refno by1}
\def\cref#1{\immediate\write\rfile{#1\semi}}

\def\semi{;\hfil\noexpand\break}

\def\inputAuxIfPresent#1{\immediate\openin1=#1
\ifeof1\message{No file \auxfileName; I'll create one.
}\else\closein1\relax\input\auxfileName\fi%
}

\newif\ifWritingAuxFile
\newwrite\auxfile
\def\SetUpAuxFile{%
\xdef\auxfileName{\jobname.aux}%
\inputAuxIfPresent{\auxfileName}%
\WritingAuxFiletrue%
\immediate\openout\auxfile=\auxfileName}


\def\bye{\par\vfill\supereject%
\ifAnyCounterChanged\linemessage{
Some counters have changed.  Re-run tex to fix them up.}\fi%
\end}



\def\c{\mskip 1mu\cdot\mskip 1mu }

\def\sign{\mathop{\rm sign}\nolimits}

\def\eps{\epsilon}

\def\pol{\varepsilon}

\def\dl^#1_#2{\delta^{#1}{}_{#2}}

\def\Gbd{\dot G_B}
\def\Gbdd{\ddot G_B}

\catcode`@=11  
\def\meqalign#1{\,\vcenter{\openup1\jot\m@th
   \ialign{\strut\hfil$\displaystyle{##}$ && $\displaystyle{{}##}$\hfil
             \crcr#1\crcr}}\,}
\catcode`@=12  


\baselineskip 15pt
\overfullrule 0.5pt



\def\pol{\varepsilon}

\def\c{\,\cdot\,}

\def\spa#1.#2{\left\langle#1\,#2\right\rangle}
\def\spb#1.#2{\left[#1\,#2\right]}
\def\lor#1.#2{\left(#1\,#2\right)}
\def\sand#1.#2.#3{%
\left\langle\smash{#1}{\vphantom1}^{-}\right|{#2}%
\left|\smash{#3}{\vphantom1}^{-}\right\rangle}
\def\sandp#1.#2.#3{%
\left\langle\smash{#1}{\vphantom1}^{-}\right|{#2}%
\left|\smash{#3}{\vphantom1}^{+}\right\rangle}
\def\sandpp#1.#2.#3{%
\left\langle\smash{#1}{\vphantom1}^{+}\right|{#2}%
\left|\smash{#3}{\vphantom1}^{+}\right\rangle}
\catcode`@=11  
\def\meqalign#1{\,\vcenter{\openup1\jot\m@th
   \ialign{\strut\hfil$\displaystyle{##}$ && $\displaystyle{{}##}$\hfil
             \crcr#1\crcr}}\,}
\catcode`@=12  


\def\ref#1#2{\nref#1{#2}}
\overfullrule 0pt
\hfuzz 52pt
\hsize 6.25 truein
\vsize 8.5 truein

\loadfourteenpoint
\newcount\eqncount
\newcount\sectcount
\eqncount=0
\sectcount=0
\def\secta{\global\advance\sectcount by1
\eqncount=0}

\def\equn{
\global\advance\eqncount by1
\eqno{(\the\sectcount.\the\eqncount)}        }
\def\put#1{\global\edef#1{(\the\sectcount.\the\eqncount)}     }

\def\eps{\epsilon}
\def\pol{\varepsilon}

\def\c{\,\cdot\,}

\def\eps{\epsilon}

\def\x#1#2{x_{#1 #2}}

\def\Gbdb{\dot {\overline G}{}_B}
\def\Gbddb{\ddot {\overline G}{}_B}

\ref\Veltman{
B.S. DeWitt, Phys.\ Rev.\ 162:1239 (1967)\semi
M. Veltman, in {\it Les Houches 1975,
Methods in Field Theory}, ed R. Balian and J. Zinn-Justin,
(North Holland, Amsterdam, 1976)}

\ref\Background{G. 't Hooft,
Acta Universitatis Wratislavensis no.\
38, 12th Winter School of Theoretical Physics in Karpacz; {\it
Functional and Probabilistic Methods in Quantum Field Theory},
Vol. 1 (1975)\semi
B.S.\ DeWitt, in {\it Quantum Gravity II}, eds. C. Isham, R.\ Penrose and
D.\ Sciama (Oxford, 1981)\semi
L.F.\ Abbott, Nucl.\ Phys.\ B185:189 (1981)\semi
L.F\ Abbott, M.T.\ Grisaru and R.K.\ Schaeffer,
Nucl.\ Phys. {B229}:372 (1983)}

\ref\Short{Z. Bern and D.A.\ Kosower, Phys.\ Rev.\ Lett.\ 66:1669 (1991)}

\ref\Long{Z. Bern and D.A.\ Kosower, Nucl.\ Phys. {B379}:451 (1992)}

\ref\Pascos{Z. Bern and D.A.\ Kosower, in {\it Proceedings of the PASCOS-91
Symposium}, eds.\ P. Nath and S. Reucroft}

\ref\FiveGluon{Z. Bern, L. Dixon and D.A.\ Kosower,
Phys.\ Rev.\ Lett. 70:2677 (1993)}

\ref\Future{Z. Bern, L. Dixon and D.A.\ Kosower, in preparation}

\ref\Berends{F.A. Berends, W.T.\ Giele and H. Kuijf,
Phys. Lett.\ {211B}:91 (1988)}

\ref\KLTtree{ H. Kawai, D.C.\ Lewellen and S.-H.H.\ Tye,
Nucl.\ Phys.\ {B269}:1 (1986)}

\ref\ParkeTaylor{S.J. Parke and T.R.\ Taylor,
Phys. Rev. Lett.{56}:2459 (1986)}

\ref\Green{
M.B.\ Green, J.H.\ Schwarz and L. Brink, Nucl.\ Phys.\ B198:472 (1982)}

\ref\Scherk{ J. Scherk, Nucl.\ Phys. {B31}:222 (1971)\semi
A. Neveu and J. Scherk, Nucl.\ Phys. {B36}:155 (1972)\semi
J.\ Minahan, Nucl.\ Phys.\ B298:36 (1988)}

\ref\Mapping{Z. Bern and D.C.\ Dunbar, Nucl. Phys. {B379}:562 (1992)}

\ref\Lam{C.S.\ Lam, preprints McGill/92-32 and 92/53}

\ref\FirstQuantized{L. Brink, P. Di Vecchia and P. Howe, Phys.\ Lett.\ 65B
Nucl.\ Phys. \ B118 (1977) 76\semi
F.A.\ Berezin and M.S.\ Marinov, JETP Lett.\ 21:320 (1975)\semi
R.Casalbouni, Nouvo Cimento 33A:389 (1976)\semi
M.B.\ Halpern and  P. Senjanovic, Phys.\ Rev.\ D15:1655 (1977)\semi
M.B.\ Halpern, A. Jevicki and P. Senjanovic, Phys.\ Rev.\ D16:2476 (1977)\semi
E.S.\ Fradkin and A.A.\ Tseytlin, Phys.\ Lett.\ 158B:316 (1985);
163B:123 (1985); Nucl.\ Phys.\ B261:1 (1985)}

\ref\Strassler{M. Strassler, Nucl.\ Phys.\ {B385}:145 (1992)}

\ref\Ven{A. van de Ven, Nucl.\ Phys.\ {B378}:309 (1992)}

\ref\Bosonic{Z. Bern, Phys.\ Lett.\ 296B:85 (1992)}

\ref\Tasi{Z. Bern, UCLA/93/TEP/5, hep-ph/9304249, procceedings of TASI 1992}

\ref\Cho{H.T.\ Cho, K.L.\ Ng, Phys.\ Rev.\ D47:1692 (1993)}

\ref\XZC{%
F.\ A.\ Berends, R.\ Kleiss, P.\ De Causmaecker, R.\ Gastmans and T.\ T.\ Wu,
        Phys.\ Lett.\ 103B:124 (1981)\semi
P.\ De Causmaeker, R.\ Gastmans,  W.\ Troost and  T.\ T.\ Wu,
Nucl. Phys. B206:53 (1982)\semi
R.\ Kleiss and W.\ J.\ Stirling,
   Nucl.\ Phys.\ B262:235 (1985)\semi
   J.\ F.\ Gunion and Z.\ Kunszt, Phys.\ Lett.\ 161B:333 (1985)\semi
 R.\ Gastmans and T.T.\ Wu,
{\it The Ubiquitous Photon: Helicity Method for QED and QCD} (Clarendon Press)
(1990)\semi
Z.\ Xu, D.-H.\ Zhang and L. Chang, Nucl.\ Phys.\ B291:392 (1987)}

\ref\Susy{ M.T.\ Grisaru, H.N.\ Pendleton and P.\  van Nieuwenhuizen,
Phys. Rev. {D15}:996 (1977) \semi
M.L.\ Mangano and S.J. Parke, Phys.\ Rep.\ {200}:301 (1991)}

\hfill IASSNS-HEP-93-31

\hfill SWAT-93-01

\hfill May  1993

\vskip - 2 cm

\Title{\vbox{\centerline{String-Based Methods in
Perturbative Gravity}} }

\vskip .3 cm
\centerline{\bf Zvi Bern\footnote{*}{Permanent address: Dept.\ of Physics,
UCLA, LA, CA 90024}}
\centerline{\it Institute for Advanced Study}
\centerline{\it Olden Lane}
\centerline{\it Princeton, NJ 08540}
\vskip .3 cm
\centerline{\bf David C. Dunbar }
\centerline{\it Department of Physics}
\centerline{\it University College of Swansea}
\centerline{\it Swansea SA2 8PP, U.K.}
\smallskip \centerline{\rm and} \smallskip
\centerline{\bf Tokuzo Shimada }
\centerline{\it Natural Science Division}
\centerline{\it Izumi Campus, Meiji University}
\centerline{\it Eifuku 1-9-1, Suginami, Tokyo 168, Japan}

\vskip 1.2 truecm \baselineskip12pt

\vskip .5 cm
\centerline{\bf Abstract }

\vskip 0.3 truecm
{
\narrower\smallskip
String theory implies a relatively modest growth in computational
complexity for perturbative gravity calculations as compared to gauge
theory calculations, contrary to field theory expectations.  An
explicit string-based calculation, which would be extremely difficult
using conventional techniques, is presented to illustrate this.
\smallskip}

\baselineskip14pt

\vfill\break


\vskip .4 truecm
\secta
\noindent
{\it \the\sectcount . Introduction.}
Perturbative computations in gravity are notorious for their algebraic
complexity, being many orders of magnitude more complicated than the
corresponding gauge theory computations.  For example, a brute force
computation of the one-loop four-graviton scattering amplitude using
conventional Feynman diagram techniques [\use\Veltman] involves
$\sim 10^8$
terms. Even with the background field method [\use\Background] in a
brute force computation one would encounter $\sim 10^6$ terms.  The
size of these intermediate expressions may be compared to the final
results which are quite compact; indeed the amplitude for one minus
and three plus helicities fits on a line.

Recently, a string-based technique significantly more efficient
than conventional Feynman diagram techniques was developed for the
computation of one-loop $n$ gluon amplitudes
[\use\Short-\use\Future].  Obvious questions are whether this
technique can be extended to other cases and whether string theory
provides additional non-trivial guidance for these extensions.  In
this letter, we address these questions by extending these
string-based techniques to perturbative gravity.
At tree level, Berends, Giele and Kuijf [\use\Berends]
have already used string theory [\use\KLTtree] to give compact expressions
for a class of tree-level gravity amplitudes using known Yang-Mills
tree amplitudes [\use\ParkeTaylor].
For $N\!=\!8$ supergravity, Green, Schwarz and Brink have used
the Green-Schwarz formulation of
string theory to give compact results for the four
graviton one-loop amplitude [\Green].

The string-based technique was originally developed to compute
one-loop gluon matrix-elements that are formidable to compute but
which are required for current and future experiments.  This
technology was a key ingredient in the first calculation of the
one-loop five-gluon amplitude (which will enter into the analysis of
three-jet events at hadron colliders) [\use\FiveGluon].  This
technique has been summarized in terms of systematic rules
[\use\Long,\use\Pascos] for the one-loop $n$-gluon amplitude which
require no knowledge of string theory and bypass much of the algebra
associated with Feynman diagram calculations.

To convert the rules to the case of one-loop graviton scattering
amplitudes we alter the details of the string construction
to recover gravity amplitudes rather than gauge theory
amplitudes in the infinite string tension limit [\use\Scherk].
Since
the string based rules for gauge theories are already computationally
efficient one expects considerable advantages in using string-based rules
for gravity.

Calculations of one-loop gravity amplitudes have never been performed
using traditional Feynman diagram methods.  With the string-based
method [\use\Long,\use\Pascos] we exhibit a four-point
graviton-by-graviton scattering calculation for a particular helicity
configuration and arbitrary particle content.
This computation would be exceedingly difficult by
traditional Feynman diagram techniques but is very simple with string
based techniques.

Given the conventional field theory understanding of the efficiency of
the string-based
methods when performing a one-loop Yang-Mills
calculation [\use\Mapping], one might think to apply this knowledge to
other cases without further appeals to string theory [\use\Lam].
However, for the case of one-loop gravity the field theory
understanding of Yang-Mills is insufficient to obtain the full benefit of the
string-based methods.  In particular, the
structure of the string integrand is
$$
(\hbox{Closed String}) \sim (\hbox{Open String})^2 \; .
$$
Since closed strings contain gravity and open strings contain gauge theory
there should be a formulation of gravity with the property that the integrands
of the diagrams satisfy
$$
(\hbox{Gravity}) \sim (\hbox{Yang-Mills})^2 \; .
\eqn\Gravity
$$
In string theory this relationship can be made precise.

Given that string theory has this property one can attempt to
reorganize field theory to mimic this.  To do so non-trivial
field redefinitions and gauge choices are required.
In this way one can attempt to mimic the string simplicity
of the amplitude,
but in a conventional field theory approach there is no guiding
principle.  (The field theory first quantized
formalism [\use\FirstQuantized,\use\Strassler] could be used for
studying the effective action.)

\vskip .4 truecm
\secta
\noindent
{\it \the\sectcount . Field theory structure.}
We will now examine the properties that a
reorganization of conventional field theory must satisfy
to mimic the string-based structure (\Gravity).
The starting point in field theory
is the Einstein-Hilbert action
$$
S[h] = {2 \over \kappa^2} \int d^4 x \, \sqrt{-g} R \; .
$$
Our conventions
are chosen so that the kinetic term has the correct canonical normalization.
The metric is expanded as $g_{\mu\nu} = \eta_{\mu\nu} + \kappa h_{\mu\nu}$
where $h_{\mu\nu}$ is the graviton field.
The first step in finding a conventional
field theory formulation which mimics string theory is to find a suitable
propagator for $h_{\mu\nu}$.
In string theory the propagator $(L_0 + \tilde L_0 - 2)^{-1}$,
where $L_0$ and $\tilde L_0$ are left- and right-mover world-sheet
Hamiltonians,  does
not contain Lorentz indices.  This indicates that the required field
theory propagator should have a trivial Lorentz structure
and therefore be proportional to the unit tensor
$$
I_{\mu\nu; \rho\sigma} = {1\over 2}
\Bigl(\eta_{\mu\rho}\eta_{\nu\sigma} + \eta_{\mu\sigma}\eta_{\nu\rho} \Bigr)
\; .
$$
The unit tensor is
a symmetrization of a product of $\eta_{\mu\nu}$'s which is the tensor
in the propagator of Feynman-like gauges in Yang-Mills.
The commonly used de Donder gauge gravity propagator [\use\Veltman]
is
$$
P_{\mu\nu; \rho\sigma} =
{i } \Bigl[ { I_{\mu\nu;\rho\sigma}\over  p^2 + i \varepsilon}
 - {1 \over D-2} {\eta_{\mu\nu} \eta_{\rho\sigma}
       \over p^2 + i \varepsilon} \Bigr]
$$
where the signature of $g_{\mu\nu}$ is $(+,-,-,-)$.
This propagator is not of the desired form, since there is an
extra trace piece and so the de Donder gauge is not
an appropriate candidate to mimic the string organization.  Although
it is not possible to obtain a
propagator with only a unit tensor within the class of standard gauges,
since the de Donder gauge propagator is close to the desired form one
might suspect that there exists a modification of the theory
with the desired field theory propagator.
String theory suggests a natural way of accomplishing this.

In string theory there is always an additional field
associated with the graviton -- the dilaton $\varphi$.  This suggests that
one can add a dilaton to the theory in order to produce a simple propagator
to aid in calculations.  At the end of a calculation one would subtract out
the dilaton contribution, which is quite simple because it
is a scalar.
In a string-based calculation one also needs to
subtract the dilaton contribution.
In string theory there is in addition an
antisymmetric tensor which must be subtracted;
in four dimensions, this is effectively another
scalar.

{}From the field theory understanding of the gauge theory rules
[\use\Mapping], the
background field method [\use\Background]
is needed to mimic the loop part of the
string-based rules.
Consider the one-loop effective action of gravity coupled to a dilaton
and carry out a background field
expansion $g_{\mu\nu}=\bar{g}_{\mu\nu}+\kappa h_{\mu\nu}$.
With the background field de Donder gauge choice, the
part of the action
quadratic in the quantum fields is
$$
S =
\int d^4 x \sqrt{-\bar g} \Bigl[-{1\over 2} h_{\mu\nu} D^2 h^{\mu\nu}
+  h^{\mu\nu} R_{\mu\rho \nu\sigma} h^{\rho\sigma}
+ {1\over 4} h^{\mu}_{\mu} D^2 h^{\mu}_{\mu} - {1\over2} \varphi D^2 \varphi
- \bar \chi^\mu D^2 \chi_\mu \Bigr]
$$
where we have used the on-shell conditions on the background field
and have included the
ghosts $\chi_\mu$. The curvature and covariant
derivatives are with respect to the background field.
Consider the field redefinition
$$
\eqalign{
h_{\mu\nu}  = \tilde h_{\mu\nu} +
{ \eta_{\mu\nu}
\over \sqrt{D-2} }
\tilde \varphi
\hskip 0.5 truecm ;  \hskip 1.5 truecm
\varphi  = {1\over \sqrt{2} } \tilde h^{\mu}_{\mu}
+ \sqrt{{D - 2\over 2} } \tilde\varphi \; . \cr}
\eqn\Redefin
$$
This has no effect on the value of the effective
action since it is only a change of variables for the internal quantum field.
(There is a trivial Jacobian in the path integral which
is unity in dimensional regularization.)
Performing the field redefinition yields
$$
S = \int d^4 x \sqrt{-\bar g} \Bigl[ -{1\over 2} \tilde
h_{\mu\nu} D^2 \tilde h^{\mu\nu}
+  \tilde h^{\mu\nu} R_{\mu\rho \nu\sigma} \tilde h^{\rho\sigma}
+{1\over 2}
\tilde\varphi D^2 \tilde\varphi - \bar \chi^\mu D^2 \chi_\mu\Bigr]
$$
where again we have dropped terms that vanish after imposing the
equation of motion on the background field $\bar{g}$.  In this action
the `graviton' propagator is proportional to the
unit tensor and is thus of the required
form to mimic the string organization.
Furthermore, the background field graviton three vertex $G_3$
derived from this action is
$$
\eqalign{
G_3{}_{\mu\nu\rho}^{\kappa\lambda\delta}(k, p , q) & = -{i \over 8} \kappa
\Bigl[
V_3{}_{\mu\nu\rho}(k, p, q) \times V_{3}{}^{\kappa\lambda\delta}(k, p, q)
 + \{\mu \leftrightarrow \kappa\} , \{\nu\leftrightarrow\lambda \},
\{\rho\leftrightarrow\delta \} \Bigr] \cr }
$$
where $V_3$ is proportional to the kinematic
part of the Feynman gauge background field  Yang-Mills
three-vertex,
$$
V_3{}_{\mu\nu\rho}(k, p , q) = \eta_{\nu\rho} {(p - q)_\mu \over 2}
- \eta_{\mu\rho} k_\nu +  \eta_{\mu\nu}k_\rho
$$
where $k$ is the momentum of the on-shell background field line and $p$ and
$q$ are the momenta of the internal lines.  The vertex $G_3$ is therefore
of the desired form to mimic string theory since it is composed of
products of Yang-Mills vertices.

In background field method one would sew tree diagrams in some other
gauge onto the one particle irreducible diagrams [\use\Background].
For the tree parts of diagrams, the relevant gauge choices and field
redefinitions necessary to mimic the string form are more complicated
but are similar (although not identical) to the choices made by van de
Ven [\use\Ven] in his computation of the two-loop infinity of gravity.
A field redefinition is also needed
to remove the trace term in the tree-level graviton propagator.

One can expect that this process of reformulating field theory to
mimic the string-based structure can be continued, but the process
becomes increasingly obscure. For example, Yang-Mills only has three-
and four-point vertices while gravity has infinitely many vertices.%
\footnote{*}{In the field theory limit of string theory
higher point vertices appear from a combination of $\delta$-functions
in the Schwinger proper time and by cancellation of kinematic poles
against factors in the kinematic expression.}
A simpler approach to carry out calculations is to proceed
directly using string theory.  The procedure for obtaining field theory
rules from string theory has been described in
refs.~[\use\Long,\use\Bosonic].

\vskip .3 truecm
\secta
\noindent
{\it\the\sectcount . One-loop rules for gravity.}
The one-loop string-based rules for gravity are similar to those
for gauge theory [\use\Long] so we only outline the differences
between the two sets of rules.  We use the bosonic string form of
the rules [\use\Bosonic,\use\Tasi] since the kinematic expression is
simpler than the heterotic string form originally used
[\use\Long,\use\Pascos] although it contains identical information.
(The heterotic string was used in the original derivation of the rules
because of its full consistency.)

The starting point of these rules are labeled $\phi^3$ diagrams
(excluding tadpoles.)  Considering that gravity has an infinite set of
higher point Feynman vertices, a description in terms of $\phi^3$
diagrams may seem surprising, but the contributions from all such higher
vertices are implicitly included.  There is no need to consider
diagrams with loops isolated on external legs as these vanish in
dimensional regularization.

The external legs of the diagrams should be labeled in the same way as
ordinary Feynman diagrams with all orderings included.  This
is unlike the gauge theory case where the legs were color ordered.
The inner lines of
a tree attached to a loop are labeled according to the rule that as
one moves from the outer lines to the inner lines, one labels the
inner line by the label of the most clockwise of the two outer lines.
(See refs.~[\use\Long,\use\Pascos,\use\Tasi] for further details.)
According to the rules, each labeled $n$-point
$\phi^3$-like diagram evaluates to
$$
\eqalign{
{\cal D} =
i { (-\kappa)^n \over (4\pi)^{2-\eps/2} }
 \Gamma(n_\ell-2+\eps/2)
&\int_0^1 dx_{i_{n_\ell-1}} \int_0^{x_{i_{n_\ell-1}}} dx_{i_{n_\ell-2}} \cdots
\int_0^{x_{i_3}}  dx_{i_2} \int_0^{x_{i_2}} dx_{i_1} \cr
& \times
{K^{}_{\rm red} \over
\Bigl(\sum_{l<m}^{n_\ell} P_{i_l}\c P_{i_m} \x {i_m}{i_l}
(1-\x {i_m}{i_l})\Bigr)^{n_\ell -2  +\eps/2}}  \cr}
\eqn\IntegrationRule
$$
where the ordering of the loop parameter integrals corresponds to the
ordering of the $n_\ell$ lines attached to the loop, $x_{ij} \equiv x_i
- x_j$, and $K_{\rm red}$ is the reduced kinematic factor.  The string-based
rules efficiently yield $K_{\rm red}$ in a compact form.
The lines attached to the loop carry momenta $P_i$ which will be
off-shell if there is a tree attached to that line.
The dimensional regularization parameter $\eps = 4 -
D$ handles all ultraviolet and infrared divergences.  The $x_{i_m}$
are related to ordinary Feynman parameters by $x_{i_m} = \sum_{j=1}^m a_j$.
The amplitude is then given by summing over all diagrams.

The starting point for evaluating $K_{\rm red}$ for any diagram
is the graviton kinematic expression
$$
\eqalign{
{\cal K} &=
\int \prod_{i=1}^n dx_i d \bar x_i \prod_{i<j}^n
\exp\biggl[ k_i\c k_j G_B^{ij} \biggr]
\exp \biggl[ (k_i\c\pol_j - k_j\c\pol_i) \, \Gbd^{ij}
         - \pol_i\c\pol_j\, \Gbdd^{ij} \biggr] \cr
& \hskip 2 cm \times
\exp \biggl[ (k_i\c\bar\pol_j - k_j\c\bar\pol_i) \, \Gbdb^{ij}
         - \bar\pol_i\c\bar\pol_j\, \Gbddb^{ij} \biggr]
\biggr|_{\rm multi-linear} \cr}
\eqn\MasterKin
$$
where the `multi-linear' indicates that only the terms linear in all
$\pol_i$ and $\bar\pol_i$ are included.  The graviton polarization
tensor is reconstructed by taking $\pol_i^\mu
\bar\pol_i^\nu \rightarrow \pol_i^{\mu\nu}$.
This kinematic expression is obtained from a bosonic string and
contains the same information as that obtained from a superstring
[\use\Bosonic].  The structure of this kinematic expression is that
the polarization factors are a product of two gauge theory factors,
[\use\Long,\use\Pascos,\use\Bosonic] corresponding to the left- and
right-movers of the underlying closed string theory.  In string theory
the $G_B$ are Green functions on the world sheet, but in the field
theory limit these become `Feynman parameter functions'.  From a
conventional Feynman diagram point of view, the existence of a
universal kinematic function is surprising as there is apparently no
simple relationship between the various Feynman diagrams contributing
to a given process.

In the form of the rules presented in refs.~[\use\Long,\use\Bosonic],
one integrates by parts to remove all $\Gbdd$. In the case of gravity the
integration by parts on the left and right are not independent and
certain cross-terms where a left-mover derivative hits right-mover terms
must be taken into account.  We will discuss these cross-terms
elsewhere since the integration by parts is not necessary for the
calculation in the next section.  It is this integration by parts
step which reduces gravity to a $\phi^3$ structure.
(This step is
not an essential part of the string-based method, which can be
formulated without integration by parts [\use\Future].)

Given the integrated by parts kinematic expression,
for a particular diagram with a two-point tree with lines labeled
by $i$ and $j$, with $i$ appearing before $j$ in the clockwise
ordering, the tree rules tell one
to replace a $(\Gbd^{ij})^n (\Gbdb^{ij})^m$ in each term by a
factor of $\delta_{n,1} \delta_{m,1} (-2 k_i \c k_j)^{-1}$.
One moves from the outside inward
iteratively, replacing the functions as described.
These tree rules do not depend
on what particles circulate in the loop and are similar to those in
refs.~[\use\Long,\use\Pascos,\use\Tasi].

After the tree rules are applied to a given diagram one then applies
loop substitution rules.
These are essentially identical rules as for Yang-Mills applied
independently to both the left- and right-mover parts of the kinematic
expression.  This provides an explicit diagram-by-diagram relationship
between the one-loop gravity amplitude and the corresponding gauge theory
amplitude.  For gravitons (and the associated ghosts) circulating in
the loop, every term generates two types of contributions.

The first contribution for left-movers
is obtained by multiplying the kinematic expression by an overall factor of
$(2 - \eps\delta_R)$ and substituting
$$
\Gbd^{ij} \longrightarrow  {1\over 2} (-\sign(\x ij ) + 2 \x ij )  \; ,
\hskip 2 cm
\eqn\BasicSubst
$$
and exactly the same substitution for the right-mover $\Gbdb^{ij}$.
The parameter $\delta_R$ depends on the precise form of the
regularization scheme used [\use\Long].
When this first type of term occurs for both left- and right-movers
instead of a factor of $(2 - \eps\delta_R)^2$, the correct factor
is $(4-\eps\delta_R)(1-\eps\delta_R)/2$, which is the number of graviton
degrees of freedom.
More generally for a theory of gravity containing various particle types,
the factor $(2 - \eps\delta_R)^2$ is replaced by
$$
N_s = N_b - N_{\!f}
$$
where $N_b$ is the number of bosonic states (including any modifications
due to dimensional regularization) and $N_{\! f}$ is the number of
fermionic states which circulate in the loop.

The second type of contribution for gravitons arises if a particular
term contains a cycle of $\Gbd$s [\use\Bosonic,\Tasi].  The rules for
cycle contributions are essentially the same as for gauge theory
except that now there are both left and right contributions.  For the
graviton in the loop one simply takes the gauge theory vector rules on
the right and on the left.

For other particles in the loop one applies rules appropriate for the
particle under consideration.  For example, a contribution from a
gravitino in the loop can be obtained by using gauge theory vector loop
rules on the left and fermion loop rules on the right.  In this way
the contribution of gravitons, gravitinos, vectors, fermions or
scalars to the one-loop gravity amplitudes can be obtained by
independently choosing gauge theory scalars, fermions, or vector loop
rules given in refs.~[\use\Bosonic,\use\Tasi] for the left and right
pieces.

Modifications to include masses for the internal fermions or scalars
is simple; the only change is in the denominator in
eq.~(\use\IntegrationRule) where the massless Feynman denominator is
replaced with one corresponding to massive states circulating in the
loop.

To illustrate the gravity rules we now turn to an explicit example.

\vskip .3 truecm
\secta
\noindent
{\it \the\sectcount . Sample calculation.}
We now calculate the ${\cal A}(1^-, 2^+, 3^+, 4^+)$ four-graviton
helicity amplitude.  From a conventional Feynman diagram point of
view, this computation requires a total of 12
distinct diagrams or 54 diagrams including permutations of external
legs.  Since gravity vertices contain many terms this would be an
extremely difficult calculation with conventional Feynman diagram
techniques; using string-based rules we show that this calculation is
in fact very easy.

The first step is to insert spinor helicity simplifications into the
kinematic expression (\use\MasterKin).  The spinor helicity method for
gravitons [\use\Berends,\use\Cho] is related to that for vectors
[\use\XZC] by
$$
\pol^{++} = \pol^+ \bar{\pol}^+ ,
\hskip 2cm \pol^{--} = \pol^- \bar{\pol}^-
$$
where $\pol^{\pm\pm}$ are the graviton helicity polarizations and
$\pol^{\pm}$ are the vector helicity polarizations defined by Xu,
Zhang and Chang.
We use the notation for spinor inner products
$\langle k_1^- | k_2^+ \rangle = \langle 1 2 \rangle$ and
$\langle k_1^+ | k_2^- \rangle = [1 2] $. Using the same choice of
spinor helicity reference momenta as in the Yang-Mills computation
of ref.~[\use\Pascos] simplifies the kinematic coefficient to
$$
\eqalign{
K= S &
(\Gbd^{13} - \Gbd^{12}) (\Gbd^{24} - \Gbd^{23})
(\Gbd^{34} + \Gbd^{23}) (\Gbd^{34} - \Gbd^{24}) \cr
\null \times &
(\Gbdb^{13} - \Gbdb^{12}) (\Gbdb^{24} - \Gbdb^{23})
(\Gbdb^{34} + \Gbdb^{23}) (\Gbdb^{34} - \Gbdb^{24}) \cr }
$$
where
$$
S =   \Bigl({s^2 t \over 4} \Bigr)^2
\Bigr({{\spb2.4}^2 \over \spb1.2 \spa2.3 \spa3.4 \spb4.1} \Bigr)^2
$$
and the Mandelstam variables are $s=2 k_1 \cdot k_2$, $t=2k_1\c k_4$
and $u = -s-t$.
Due to the special helicity configuration, $\Gbdd$s do not appear
and there is therefore no need to integrate by parts.

\def\figone{1}

The next step is to determine which diagrams vanish trivially by the
tree rules. There are a total of twelve $\phi^3$-like diagrams.
Of these, seven vanish by
the tree rules.  For example, a diagram containing a 1--4 tree
vanishes because there are no $\Gbd^{14}$ Green functions. Other
diagrams which contain a 2--3 tree vanish because the remaining
factors vanish after setting the labels of the two pinched
legs together; in this case $(\Gbd^{34} - \Gbd^{24}) \rightarrow 0$
for $2 \rightarrow 3$.  The only non-vanishing diagrams
are the five shown in \figone a-e.

First consider \figone a.  This diagram has no trees so we immediately
apply the loop rules.  It is not difficult to check that all cycle
contributions of the loop cancel amongst themselves whether fermion
or vector rules are applied to the right- or left-movers.  Thus the reduced
kinematic expression can be obtained by applying the substitution rule
(\use\BasicSubst) and multiplying by the number of states $N_s$
yielding the Feynman parameter polynomial
$$
N_s S \, x_2^2 (1-x_3)^2 (x_3 - x_2)^4 \; .
$$
Up to an overall constant this is {\it precisely}
the square of the Yang-Mills Feynman parameter polynomial for the
corresponding diagram derived in refs.~[\use\Pascos,\use\Tasi].
Inserting this into the loop integral yields
$$
D_a = { i\kappa^4  \over (4\pi)^2 }
 N_s S
\int_0^1 d x_3 \int _0^{x_3} d x_2 \int_0^{x_2} d x_1 \;
{x_2^2 (1-x_3)^2 (x_3 - x_2)^4 \over [s x_1 (x_3 - x_2)
+ t (x_2 - x_1)(1-x_3) ]^2} \; .
$$
Since this and all following integrals
are finite
we have set the dimensional regularization parameter $\eps$ to zero.
This integral is easy to evaluate as the numerator
cancels the denominator after performing the $x_1$ integral.
Diagrams \figone b and \figone c are just as easy to evaluate.
The three contributions are
$$
D_a =  {i\kappa^4\over (4\pi)^2}   {N_s S \over 840 st}
, \; \hskip 1.0 truecm
D_b =  {i\kappa^4\over (4\pi)^2}  { N_s S \over 840 ut}
, \; \hskip 1.0 truecm
D_c  = {i\kappa^4 \over (4\pi)^2}  { N_s S \over 252 su} \; .
$$
This takes care of the box diagrams.

Now we evaluate the two triangle diagrams.
First consider \figone d.
Applying the rules for a 1--2 tree
reduces the kinematic coefficient (\use\MasterKin) to
$$
\eqalign{
K= -{1\over s} S
 (\Gbd^{24} - \Gbd^{23})
(\Gbd^{34} + \Gbd^{23}) (\Gbd^{34} - \Gbd^{24})
(\Gbdb^{24} - \Gbdb^{23})
(\Gbdb^{34} + \Gbdb^{23}) (\Gbdb^{34} - \Gbdb^{24}) \; . \cr }
$$
Applying the loop substitution rule yields the loop integral
$$
D_d =-i
{\kappa^4\over (4\pi)^2 }
{N_s S \over s}
\int_0^1 d x_3 \int_0^{x_3} d x_2 \;
{(1-x_3)^2 x_2^2 (x_3 - x_2)^2 \over s x_2 (x_2 - x_3) }
$$
which is a trivial integral since the denominator cancels against
the numerator.
The last non-zero diagram \figone e is similar to evaluate
and the two diagrams are
$$
D_d = {i\kappa^4 \over (4\pi)^2 } {N_s S \over 360  s^2} , \hskip 2 truecm
D_e = {i\kappa^4 \over (4\pi)^2 } {N_s S \over 360 u^2} \; .
$$

Summing over all diagrams we have the four-graviton amplitude in
a theory with any particle content as
$$
{\cal A} (1^-, 2^+, 3^+, 4^+) =
{i\kappa^4  \over (4\pi)^2 }  {N_s \over 5760}
{s^2 t^2 \over u^2} (u^2 - s t)
\Bigr({{\spb2.4}^2 \over \spb1.2 \spa2.3 \spa3.4 \spb4.1} \Bigr)^2 \; .
$$
For pure gravity $N_s =2$ because the graviton has two helicity states.
It is easy to verify that this amplitude has the required crossing
symmetry under the interchange of legs 2, 3 and 4.  In a supergravity
theory with equal numbers of bosonic and fermionic states $N_s = 0$ so the
amplitude vanishes in agreement with the supersymmetry identities
[\use\Susy].  Note that in this string-based calculation this identity
holds at the level of the integrand.

The helicity conserving process ${\cal A}(1^-, 2^-, 3^+, 4^+)$ is more
difficult to compute since cycle contributions no longer vanish and
it is infrared divergent.
However, even this is relatively easy to compute using the string-based
methods.

These calculations may be compared to the corresponding QED
calculation of light-by-light scattering.  One has a few more diagrams
and more complicated Feynman parameter polynomials to integrate, but
the extra complication is very slight when compared to the
traditional field theory expectation that gravity computations are exceedingly
more complicated than QED ones.

In conclusion, gravity provides a further example of how string-based methods
can be used to obtain results which would be extremely difficult to
obtain using field theory methods.  We expect new methods based on
string theory to have further non-trivial applications to field theory
calculations.

We thank Lance Dixon and David Kosower for helpful discussions.
This work was supported by a S.E.R.C. advanced fellowship, NATO grant
CRG-910285 and by the Texas National Research Commission grant
FCFY9202.

\vfill\eject\immediate\closeout\rfile
\centerline{{\bf References}}\bigskip\frenchspacing%
\input refs.tmp\vfill\eject\nonfrenchspacing

\break
\centerline{\bf Figure Caption}

\vskip .5 cm
\noindent
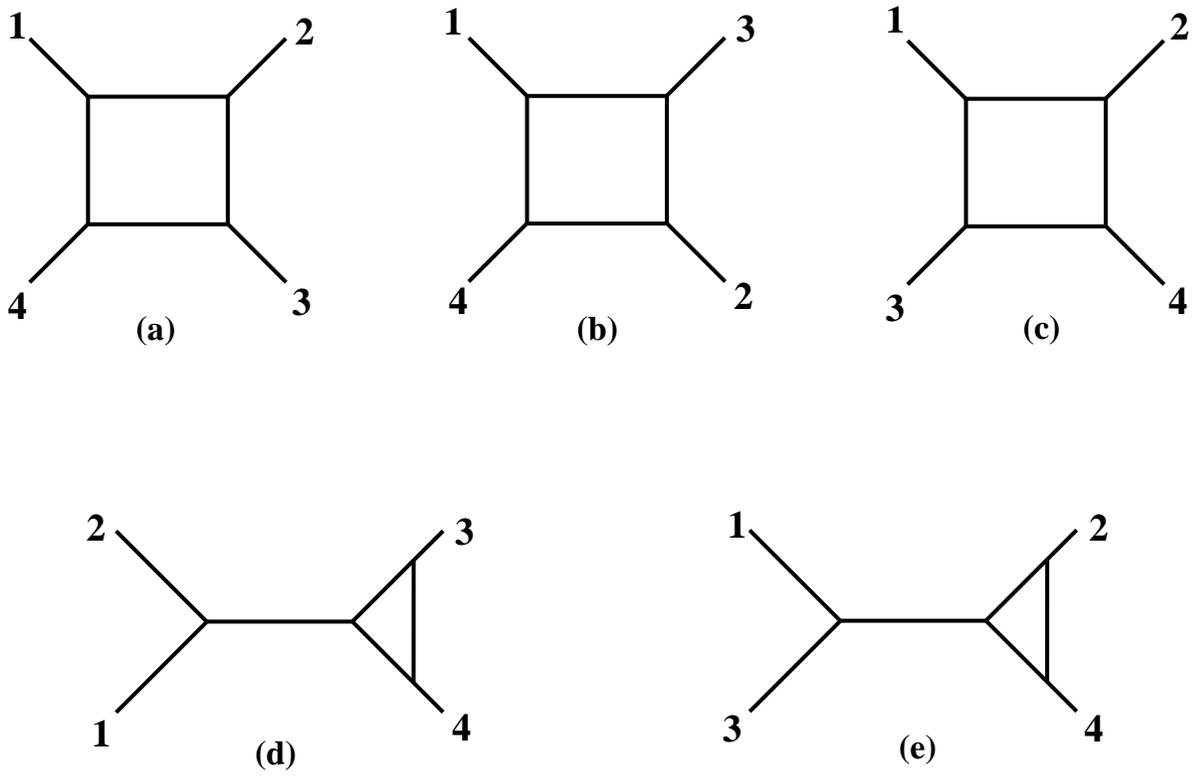
{\bf Fig 1:} The diagrams which do not vanish after applying the tree rules.

\bye